# Digital Wire Analyzer of Mechanical Tension, Electrical Continuity, and Isolation

Sebastien Prince, Pratyush Anand, James Battat, Russell Farnsworth, Nathan Felt, Roxanne Guenette, Shion Kubota, Austin Li, Em Murdock, John Oliver, Chris Stanford, and Jackson Weaver

*Abstract*—A digital instrument that allows the measurement of the mechanical tension of an array of wires of known length and density, and the testing of their electrical continuity and isolation, has been developed. The instrument measures wire tension by measuring the fundamental frequency of the wire. Its working principle is to apply direct high voltages on neighboring wires of a wire under test and sweeping the frequency of an alternating high voltage that is also applied on those neighbors. A resonance is observed in the readout signal of the middle wire when the frequency of the alternating high voltage coincides with its fundamental frequency. The instrument automates the process over 128 wires, with eight read out simultaneously. An accuracy of 1% in the measurement of tension is achieved by this digital wire analyzer (DWA).

## I. INTRODUCTION

WIRES are a common readout part of detectors used in particle physics due to their sensitivity to the presence of ionization electrons. For the proper operation of such detectors, some wire properties need to be controlled. For example, the mechanical tension needs to be at a specific value. Too high of a tension could cause the wire to break or for the structure holding the wire to deform. Too low of a tension could either cause readout noise or make the wire touch other parts of the detector, compromising it. In addition, the electrical path the wire creates needs to be continuous from the front-end electronics to the full length of the wire. Electrical discontinuity can occur if a wire breaks, if it is badly soldered onto the readout path, or if there's a discontinuity in the readout path itself, such as a broken trace in a printed circuit board (PCB). Also, the wire needs to be electrically isolated from its environment to be sensitive to the ionization electrons.

The latter two properties can usually be tested using off-the-shelf electronic instruments. The measurement of the mechanical tension does not have such ready solutions however. As such, different techniques have been developed. Since the wire length and density are known in particle detectors, the tension can be determined by measuring either the wire sagitta under deflection [1] or the fundamental frequency of the wire by exciting it. This latter approach can further be categorized by the way the excitation is achieved: via a mechanical force [2], [3], a magnetic force [4], [5], [6], [7], [8], or an electrical force [9], [10], [11], [12], [13], [14], [15], [16], [17], [18]. All these methods achieve a measurement accuracy of 1% or better, with the exception of [18] in which a value of 3% is reported.

Manuscript received 12 June 2022; revised 12 September 2022; accepted 7 October 2022. Date of publication 31 October 2022; date of current version 14 November 2022. This work was supported in part by Fermi Research Alliance, LLC, under Contract 665148 and in part by the U.S. Department of Energy, Office of Science, Office of High Energy Physics, under Award DE-SC0007881. The work of Sebastien Prince was supported in part by the Natural Sciences and Engineering Research Council of Canada (NSERC) under Grant 516912 and in part by the Fonds de recherche du Québec—Nature et Technologies (FRQNT) under Grant 257887 and Grant 296042. The Associate Editor coordinating the review process was Anoop Chandrika Sreekantan. *(Corresponding author: Sebastien Prince.)*

Sebastien Prince, Nathan Felt, Shion Kubota, John Oliver, and Chris Stanford are with the Department of Physics, Harvard University, Cambridge, MA 02138 USA (e-mail: sprince@fas.harvard.edu; felt@g.harvard.edu; skubota@g.harvard.edu; john.nathaniel.oliver@gmail.com; cstanford@g.harvard.edu).

Pratyush Anand was with the Department of Physics, Harvard University, Cambridge, MA 02138 USA. He is now with the Department of Physics, ETH Zurich, 8093 Zurich, Switzerland (e-mail: panand@student.ethz.ch).

James Battat is with the Department of Physics, Wellesley College, Wellesley, MA 02481 USA (e-mail: jbattat@wellesley.edu).

Russell Farnsworth, Austin Li, and Em Murdock are with the Harvard College, Harvard University, Cambridge, MA 02138 USA (e-mail: rsfarnsworth@college.harvard.edu; awli@college.harvard.edu; emilymurdock@college.harvard.edu).

Roxanne Guenette was with the Department of Physics, Harvard University, Cambridge, MA 02138 USA. She is now with the Department of Physics and Astronomy, University of Manchester, M13 9PL Manchester, U.K. (e-mail: roxanne.guenette@manchester.ac.uk).

Jackson Weaver was with the Harvard College, Harvard University, Cambridge, MA 02138 USA. He is now with the Department of Materials Science and Metallurgy, University of Cambridge, CB3 0FS Cambridge, U.K. (e-mail: jacksonweaver@college.harvard.edu).

Digital Object Identifier 10.1109/TIM.2022.3214606

The control of wire properties is needed to build detectors such as those in the upcoming DUNE experiment [19]. DUNE is an international experiment that aims to measure a fundamental physical constant for the first time, detect neutrinos produced in supernovas, and possibly discover new physical interactions altogether. One of its detector modules contains close to one million wires, which are assembled in large arrays of equally spaced parallel wires in multiple layers [20]. These arrays are themselves subdivided into readout groups of 128 wires. Since the measurement methods based on deflection or on mechanical or magnetic excitation require physical access to the wires, or having the wire inside a magnetic field, they are not well-suited for such detectors. On the other hand, the electrical excitation methods are appropriate.

The electrical methods differ among themselves mainly in how the electric force is applied, which depends on the detector geometry. In the case of the open-frame wire arrays of DUNE, a method that is particularly well-suited for them is that of Garcia-Gamez et al. [18], which makes use of direct and alternating high voltages applied on the wires themselves. Indeed, it does not require conductive electrodes near the wires, the high voltages can be safely applied onto wires that have a pitch of a few millimeters, and it is scalable to a large number of wires. An instrument based on that method that automates the measurement process on a large number of wires could then be used for the DUNE detector module. Furthermore, since such an instrument would apply and read out voltages on multiple wires, it could also be used to test electrical continuity and electrical isolation.

Such a digital wire analyzer (DWA) has been designed, produced, and tested. It consists of custom hardware, firmware, and software that can connect to 128 wires and read out eight of them simultaneously. It can be seen as an extension to the work that has been done in [18]. The main improvements are an increased number of connected channels, an ability to switch the readout automatically among the connected channels, an improved noise suppression, an independence on external electronics other than a computer, and a capability to test electrical continuity and isolation. These improvements allow for automating the measurement process, which significantly reduces the amount of manipulation needed to analyze the properties of wires in large arrays, thus making the process faster and more convenient, and for improving the accuracy and frequency range of the measurement.

A summary of the method for measuring wire tension on which the DWA is based is now given, followed by descriptions of the custom hardware, firmware, and software. These descriptions are followed by some illustrative results that show the capabilities of the instrument.

## II. Method

When the linear density $\lambda$, length $L$, and fundamental frequency $f_0$ of a wire are known, its tension $T$ can be calculated as

$$T = 4\lambda L^2 f_0^2. \tag{1}$$

Therefore, the measurement of tension becomes a measurement of fundamental frequency for arrays of wires with known length and density.

The method used by the DWA to measure $f_0$ relies on the wire under test having two equally spaced and parallel neighboring wires in a single plane [18]. Excitation of the middle wire is achieved by applying a combination of alternating current (ac) and direct current (dc) at high voltage on its neighboring wires. Specifically, stimulus voltages $V_+(t)$, applied on one of the neighbor wires, and $V_-(t)$, applied on the other neighbor wire, are

$$V_\pm(t) = V_{ac} \sin \omega t \pm V_{dc} \tag{2}$$

where $\omega = 2\pi f$, with $f$ being the frequency of the ac voltage. The dc voltages of opposite polarity on the neighboring wires create an electric field that surrounds the wire under test. The ac voltage present on both the neighbor wires capacitively charges the middle wire. This combination results in the middle wire being driven by a sinusoidal force at the frequency of the ac voltage. The oscillatory motion of the middle wire changes the capacitance of the wire system, which gives rise to a bipolar resonance in the amplitude of the current in the middle wire for frequencies of the ac voltage around $f_0$. The measurement of the fundamental frequency of a wire under test is then a matter of sweeping the stimulus frequency and finding the center of this bipolar resonance on top of a capacitively rising baseline. More details on the bipolar resonance and the baseline are given in the Appendix.

In this measurement method, three wires are needed. The method can be scaled to a larger number of wires by reusing wires with stimulus voltages to excite both their neighbors, and reading out both, in a pattern that can be repeated. The smallest unit of this wire pattern is composed of these four consecutive wires: a wire with voltage $V_+$, a wire that is measured, a wire with voltage $V_-$, and a second measured wire. To measure all the wires, the stimulus wires need to swap roles with the measured wires.

The described measurement method of wire tension requires that the stimulus and readout wires be electrically continuous and isolated. Reciprocally, the stimulus wires can be used to test the electrical continuity and isolation of the readout wires. Since the current in the wire under test is proportional to the capacitance, which is itself proportional to the wire length, a change in the length of the electrical path results in an observable change in the amplitude of the current. An advantage of this approach compared with using a dedicated instrument such as a multimeter is that the continuity can be tested even if there is no physical access to the wire. A test of the isolation of the wires among their neighbors is also performed since, if the wire under test is not isolated from one of its neighbors, it will be at the same voltage as that stimulus wire.

## III. Hardware

Custom hardware for DWA, fulfilling the requirements of the measurement method, has been developed. It can connect to 128 wires and measure eight of them at a time. Its only external connections are to a wall outlet for mains power, to a computer network for data processing, and to the 128 wires to be measured. This number of wires is chosen such as to suit the needs of the DUNE wire arrays, but the instrument can be adapted to interface to a smaller number of wires. The DWA is composed of one commercial system-on-module (SOM) PCB running custom firmware and 13 custom PCBs: one power distribution board, one analog board, two relay boards, and nine bandpass filter (BPF) boards. An outline of these boards and their main components is shown in Fig. 1. Detailed schematics and fabrication files are available online [21].

The power distribution board converts, via ac–dc converters, the mains ac voltage into separate $\pm 5$ V dc lines that power the rest of the instrument. It is user-configurable for either the 120- or 230-V ac mains standard. This board is housed in its own enclosure, the power distribution box, to contain the electromagnetic radiation coming from the mains voltage. The power distribution box connects to the DWA box via a 20-conductor cable. The DWA box houses the other boards of the instrument in a stack, with the analog board acting as a motherboard. The SOM and BPFs connect to the analog board from the top while the relay boards are stacked underneath it.

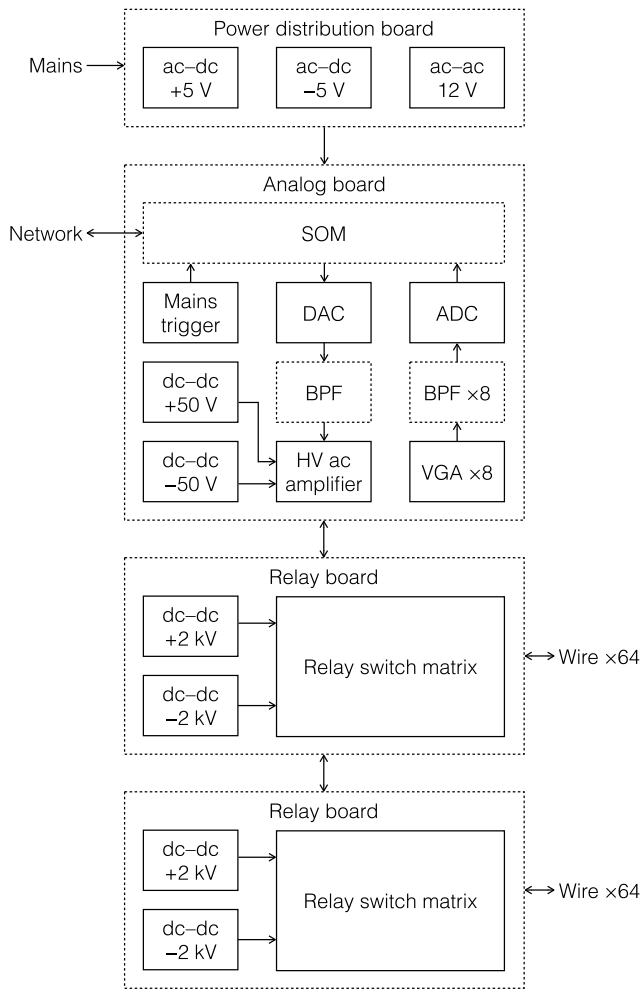

Fig. 1. Block diagram of the DWA. The dashed rectangles represent independent PCBs, while the solid rectangles represent components on the PCBs.

To produce the ac stimulus voltage, a digital-to-analog converter (DAC), part number DAC8512 from Analog Devices, is controlled to generate a square wave with a frequency corresponding to that of the ac voltage. The square wave gets filtered into its fundamental sinusoidal component by a BPF. BPF is a unity-gain eighth-order filter centered at the frequency of the ac stimulus voltage with a $Q$-factor of about 10. It is built from four cascaded second-order switched-capacitor filters provided in the Linear Technology LTC1068-200 integrated circuit. The filtering of the square wave into a sine wave results in a voltage gain of $4/\pi$ from its Fourier decomposition. The resulting sine wave is amplified by a high-voltage operational amplifier, Analog Devices ADA4700-1, with a fixed gain of 18 powered by two dc–dc converters that convert a dedicated +5 V power line into ±50 V. A high ac voltage is needed to increase the size of the capacitive signal of the wire under test relative to noise. The high-voltage sine wave is then passed to the relay boards.

The relay boards couple the ac stimulus voltage to the dc stimulus voltages of opposite polarities. These dc high voltages are obtained from two current-limited dc–dc converters. They convert a dedicated +5 V power line into voltages up to ±2 kV, configurable via potentiometers. High dc voltages are crucial as they increase the amplitude of the bipolar resonance. Each relay board contains these two converters such that the high voltages are contained within each board. The sum of the ac and dc stimulus voltages goes through a switch matrix made of high-voltage reed relays, part number SHV05-1A85-78D2K from Standex Electronics. Reed relays are chosen as switching elements due to their low capacitance of about 1 pF when open, which would otherwise compete with the capacitive signal that is measured, and their high isolation resistance in the gigaohms and high breakdown voltage in the kilovolts, which allow them to be used with the high voltages required by the measurement method. An integrated magnetic shield in the relays allows for them to be packed with high density on a PCB by preventing the magnetic fields of neighboring relays, and of other sources, from disturbing their operation. This relay switch matrix routes the stimulus voltages $V_{\pm}$ onto wires, following the pattern prescribed by the measurement method. Similarly, the same switch matrix routes the signals of the wires under test back to the analog board, where they are amplified by eight variable gain amplifiers (VGAs). Each relay board can route the stimulus voltages to, and signals from, 64 wires, for a total of 128 for the DWA. The details of the connections between the inputs and the outputs of the switch matrix and the wires are given in Fig. 2.

As shown in Fig. 2(a), the relay switch matrix consists of two consecutive sets of relays, the bus relays, and the wire relays, which are connected by 16 bus lines. Two bus relays and four wire relays are connected to each bus line. There are thus 32 bus relays and 64 wire relays for each relay board. For each bus line, one of the two bus relays routes either $V_+$ or $V_-$, while the other one routes the signal of a wire under test. Each of the four wire relays connects to a different wire, allowing to route the bus to any of these four wires. Only a single bus relay in a given bus line is active at any given time. In other words, relays of either only the left or only the right column of bus relays in Fig. 2(a) are active at a time. This connection scheme respects the repeatable pattern of four wires required by the measurement method and allows swapping the roles of wires between being a stimulus wire or a measured wire by changing which column of bus relays is active. The scheme shown in the figure repeats four times in the full switch matrix, connecting to subsequent VGAs and wires. Thus, the full scheme describes a connection to 16 wires: eight that are connected to VGAs, four that are at voltage $V_+$, and four that are at voltage $V_-$. Since the two relay boards share the eight VGA channels, bus relays connecting to a given VGA across the two relay boards are required to not be active at the same time. To allow the eight measured wires to each have two stimulus neighbor wires, the stimulus voltage of the appropriate dc polarity can be applied to a 17th wire, as shown in Fig. 2(b). Due to the repeatability of the scheme, this is done by simply activating the wire relay of that additional wire.

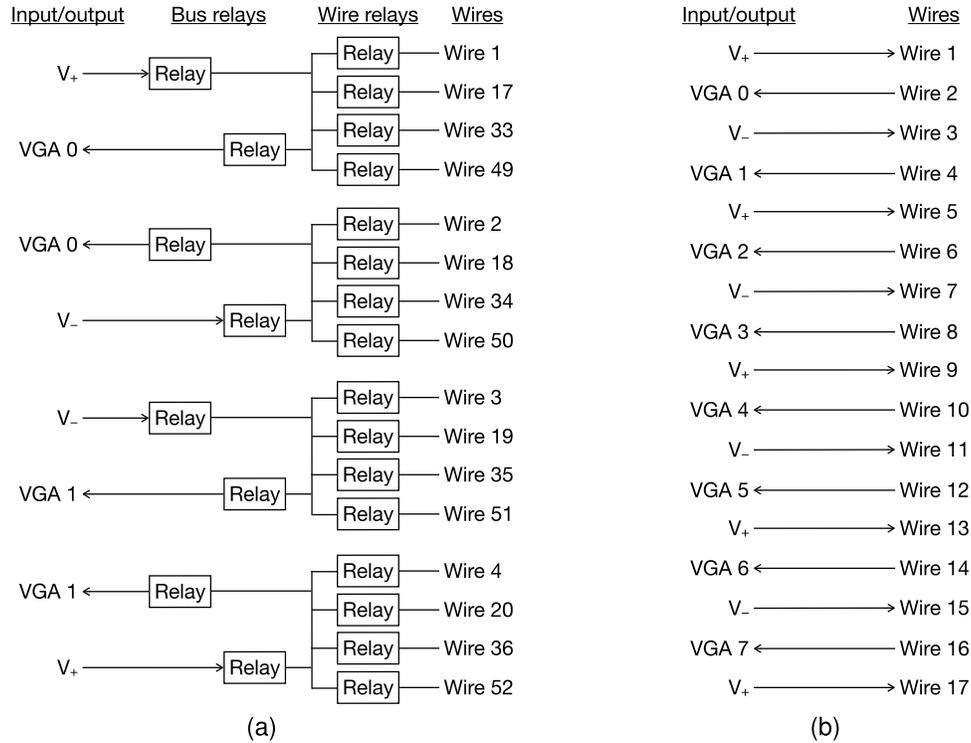

Fig. 2. Diagrams of associations between the DWA stimulus voltages and readout channels and the wires. In both the diagrams, wires with consecutive numbers are physical neighbors in an array. (a) Logical grouping of wires representing one quarter of a relay switch matrix on one relay board. (b) Physical grouping of the first 17 wires, with relays omitted for clarity.

The VGA is an operational amplifier, Analog Devices part ADA4665-2, set up as a noninverting amplifier that has a 10-k$\Omega$ resistor to ground at its inverting input. A digital potentiometer, Analog Devices AD5262, is used as feedback resistor, allowing to configure the readout gain of the VGA up to a value of 21. A resistor $R$ to ground at the noninverting input of the amplifier converts the current in the wire under test into a voltage that is read out. However, the combination of this resistor with the capacitance of the wire system creates a high-pass $RC$ filter. For a resonance to be observable in the readout voltage, $f_0$ must be located in the frequency region of the high-pass filter that sees a linearly rising gain, roughly $f_0 < 1/(2\pi RC)$. If $f_0$ is in the plateau region of the filter, roughly $f_0 > 1/(2\pi RC)$, no resonance can be observed as the filter is at unity gain and is not sensitive to small changes in capacitance. Therefore, the value of the resistor depends on the geometry of the wire array, but 100 k$\Omega$ is expected to be good for most wire configurations and is the value used.

Since a wire in the wire array acts as an antenna, its readout voltage is composed of a multitude of frequencies. The eight amplified voltages from the VGAs are each sent to a BPF to filter out noise at frequencies other than that of the stimulus voltage. These filtered signals are then sampled by an eight-channel 16-bit analog-to-digital converter (ADC) with a 4.096-V input range, Linear Technology part number LTC2320-16. The ADC counts for each channel are processed by the SOM and sent out of the DWA to the network via an Ethernet cable to be analyzed by custom computer software.

Although the BPF removes noise at frequencies different from the stimulus frequency, it does not remove noise at that frequency. This is a problem for stimulus frequencies near that of the mains voltage since the impact of that noise on the readout voltage is not negligible. To alleviate the problem, the firmware can perform an active mains noise subtraction (MNS). This technique requires a trigger signal that is obtained by stepping down the mains voltage to 12 V via an ac–ac transformer in the power distribution board, and then converting that stepped-down voltage into a square wave on the analog board.

The connection of DWA to wires is done through two polyimide printed circuits, each connecting to a different relay board, collectively referred to as the flex cable. Polyimide is chosen as the material since it provides a high dielectric strength that allows for a high density of high-voltage channels while also being mechanically flexible, allowing it to be handled as a connection cable. In addition to the flex cable, ribbon cables can be used to connect to additional stimulus wires by tapping directly into the $V_\pm$ lines on either relay board. These additional stimulus cables are necessary if there are more than 128 wires in the wire array, since the wires at either end would otherwise only be excited by a single neighbor wire. The flex cable and stimulus cables connect to an adapter board that is customized to the specific wire array under test. The flex cable and most other hardware components can be seen in Fig. 3.

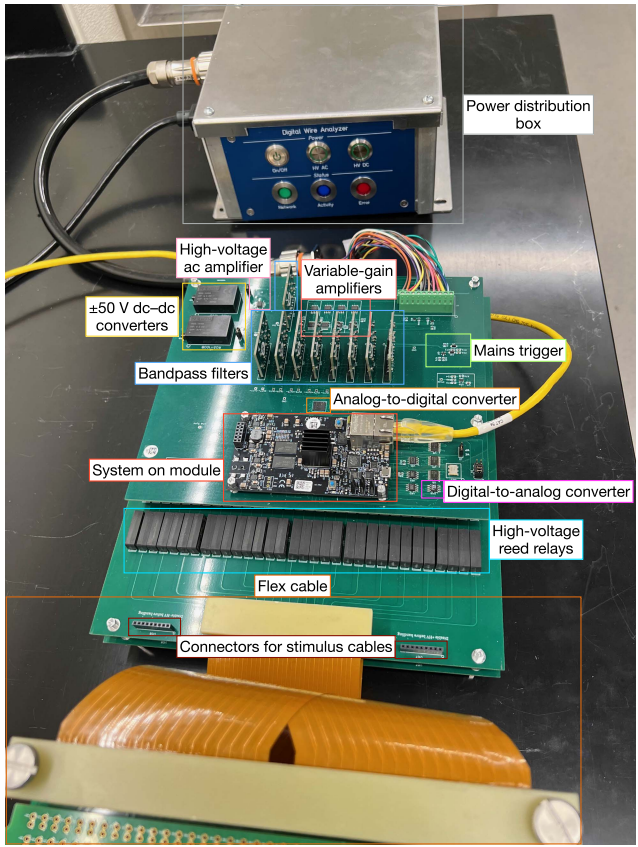

Fig. 3. Photograph of the power distribution box and of the internals of the DWA box. Colored text boxes describe components within frames of the associated color. The ±2 kV dc–dc converters and the majority of the relays on the relay boards are out of view. The wires under test are below the bottom boundary of the photograph.

TABLE I
MAIN FIRMWARE CONFIGURATION PARAMETERS

| Parameter | Value range | Resolution |
|---|---|---|
| Reed relay ×192 | Enabled or disabled | — |
| Stimulus frequency minimum | 10–1000 Hz | $\frac{1}{256}$ Hz |
| Stimulus frequency maximum | 10–1000 Hz | $\frac{1}{256}$ Hz |
| Stimulus frequency step size | $\frac{1}{256}$–990 Hz | $\frac{1}{256}$ Hz |
| Stimulus duration before sampling | 0–42.9496704 s | 2.56 µs |
| DAC output amplitude | 0–4095 mV | 1 mV |
| Digital potentiometer resistance ×8 | 60–199,278.75 Ω | 781.25 Ω |
| Number of readout ac cycles | 1–127 | 1 |
| Number of samples per ac cycle | 1–127 | 1 |
| Mains noise subtraction (MNS) | Enabled or disabled | — |
| MNS frequency minimum | 30–80 Hz | $\frac{1}{256}$ Hz |
| MNS frequency maximum | 30–80 Hz | $\frac{1}{256}$ Hz |
| ±2 kV dc–dc converters | Enabled or disabled | — |

DWA can be assigned unique media access control (MAC) and internet protocol (IP) addresses.

The firmware that runs on the FPGA carries out the automated measurement process based on the configuration it receives. The principal configuration parameters and their possible values are listed in Table I. When a frequency sweep is requested, the configured relays in the switch matrix get activated and the sweep starts at the stimulus frequency minimum and increases to the stimulus frequency maximum by stepping through the frequencies according to the stimulus frequency step size.

The FPGA allows precise control of the timing of the signals used to drive the DWA hardware. This is necessary when the stimulus frequency is high and the frequency step size is small because the BPFs require a clock frequency that is 200 times larger than their center frequency. In the extreme case of a 1-kHz stimulus frequency and a (1/256)-Hz frequency step size, the BPF clock period must change by only 19.5 ps on average. This timing requirement is met using a tapped delay line inside the FPGA.

The user can configure the stimulus duration prior to digitization. This can be necessary to allow wires to physically settle down from a resonance at a previous frequency before sampling the next one. If the duration is too short, a ringing effect can be observed in the measured voltage amplitude at the frequencies immediately following a resonance. The DAC output amplitude is used to adjust the amplitude of the ac high voltage. The readout gain of a VGA is adjusted via the resistance of its digital potentiometer. The number of ADC samples for each frequency in the sweep is determined by the product of two configuration parameters: the number of ac cycles to read out and the number of ADC samples for each of these cycles.

In the MNS technique, the contribution of the mains voltage is subtracted directly from ADC counts during a sweep. This is achieved by digitizing the mains noise at the start of the sweep, before any stimulus voltage is present and storing these ADC samples in memory. During the sweep, each ADC sample of each wire is subtracted by its associated mains noise sample. For the phase of the noise to be coherent between the two sets of samples, ADC sampling is triggered by the square wave that is generated by the analog board at the frequency of the mains voltage and in phase with it. To average out noise that is not at

## IV. FIRMWARE

The commercial SOM used in the DWA is the Avnet MicroZed based on the Xilinx Zynq-7020 system on a chip, which combines a field-programmable gate array (FPGA) with an Arm processor. The SOM has also in particular an Ethernet port and two 100-pin connectors to interface with the DWA. The custom firmware and embedded software[1] running on the SOM control all the DWA hardware components, including the DAC, BPFs, relays, and ADC, based on a configuration obtained from the network. This allows automating, via a finite-state machine, the frequency sweeps of stimulus voltages and the digitization of readout voltages.

The embedded software running on the processor implements the communication protocols used with the network and interfaces the network to the FPGA. The transmission control protocol (TCP) is used to read and write the configuration parameters from and to the SOM, while the user datagram protocol (UDP) is used to push the ADC samples out of the DWA. Due to the format used in communicating the data via UDP, the least significant bit of an ADC sample is truncated, such that only values that are multiples of two are transmitted. Multiple DWAs can be used in a single network as each the mains frequency, the readout mains voltage is oversampled by a factor of 8 compared with the number of ADC samples requested for each stimulus frequency. To reduce the number of samples stored in memory and to make the process faster, the mains noise sampling is not performed for every frequency of the sweep but rather with a 1-Hz step size. A linear interpolation across stimulus frequencies is performed for the frequencies that do not have associated noise samples. The firmware can be configured to disable the use of MNS or to perform it in a configurable frequency range.

---

[1]The source code of the firmware and embedded software can be provided upon request.

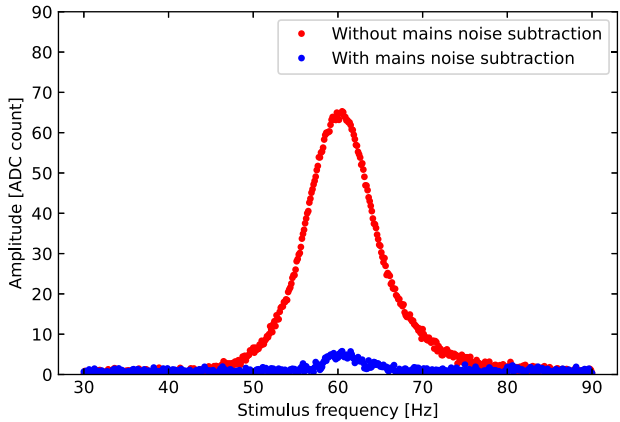

Fig. 4. Amplitude of the mains noise in the readout voltage of a wire under test as a function of stimulus frequency.

The reduction in noise due to the MNS can be seen in Fig. 4, where the MNS is performed in the 40–80-Hz range, for a mains frequency of 60 Hz. The amplitude of the mains noise is reduced by an order of magnitude with the MNS. No stimulus wires are used to produce the data in this figure, such that the shape of the noise amplitude as a function of the stimulus frequency is determined by the BPF. The level of noise obtained with the MNS at mains frequency is comparable to that caused by the second harmonic of the mains voltage and is small enough not to be detrimental to the identification of resonances in that frequency region.

The tests for electrical continuity and isolation do not require a dc high voltage as part of the stimulus voltage. As such, the FPGA can disable power to the ±2-kV dc–dc converters. Disabling dc high voltages is also needed before switching relays as otherwise a large amount of noise is seen throughout the various control lines of the SOM, which can affect the operation of the DWA.

## V. Software

Custom computer software[2] is used to configure the DWA firmware, to receive data, to identify resonances, and to record measured wire tensions. The software has a graphical user interface (GUI), based on the PyQt5 framework [22], which can be run on the Windows, Linux, and macOS operating systems. PyQtGraph [23] is used for efficient real-time data plotting and interaction.

A screenshot of the GUI is shown in Fig. 5. The GUI is separated in various tabs, each with a specific purpose and independent of the others. Also, a sidebar can be seen across all the tabs to monitor the status of the DWA at all times. The two main tabs are the Stimulus tab and the Results tab.

[2]The source code of the software can be provided upon request.

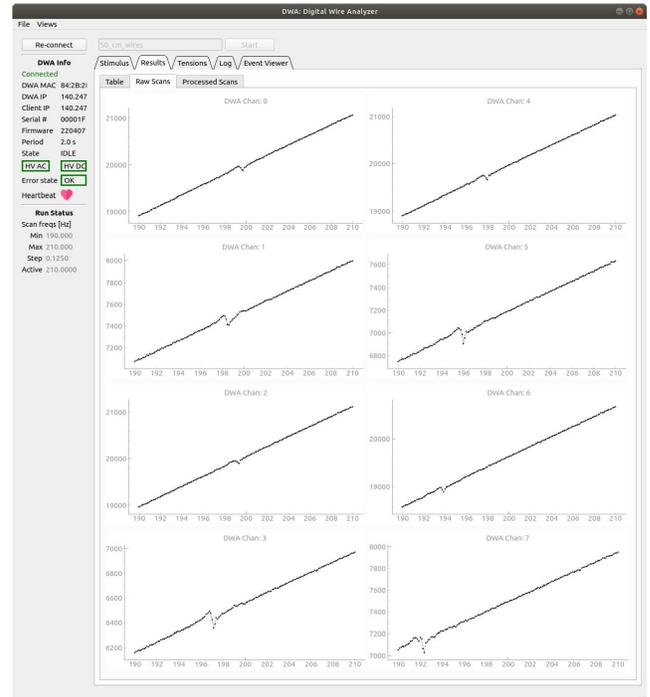

Fig. 5. Screenshot of the GUI of the DWA software running on the Ubuntu operating system. The Results tab is active, showing the voltage amplitude data of wires under test as a function of stimulus frequency for the eight DWA readout channels. Bipolar resonances can be seen in the data.

The Stimulus tab in the GUI is used to automatically configure frequency sweeps via TCP, based on knowledge specific to the wire array under test, and automate their start. Since the tests for electrical continuity and isolation do not rely on producing and resolving resonances, optimized frequency sweeps, without dc high voltage and with larger frequency step size and range, can also be configured for them. Therefore, two frequency sweeps, one optimized for the electrical tests and one for the measurement of the mechanical tension, can be performed sequentially. If the former reveals that some wires are not continuous or isolated, those wires can be omitted from the second sweep by deactivating their wire relays.

The Stimulus tab also displays in real-time the ADC samples received via UDP for each of the eight channels. Plots of voltage amplitude as a function of stimulus frequency are also displayed. The voltage amplitude is determined by performing a linear least-squares minimization of a sinusoid, with a frequency equal the stimulus frequency, to the data. The sinusoid is parameterized by a linear combination of a sine wave, a cosine wave, and a constant term, corresponding to a parameterization of the amplitude, phase, and vertical offset of the sinusoid.

The Results tab displays the results of all the frequency sweeps that have been performed on the given wire array. If a sweep tests the electrical continuity and isolation of wires, the outcome of the tests is reported for each wire. The outcome of the continuity test can be determined by comparing, on a per-channel basis, the size and shape of the amplitude values of the sweep to those obtained by the DWA when it is not connected to any wire. Similarly, the outcome of the isolation test can be determined by comparing the amplitude values of the sweep to a large threshold value.

If a sweep measures the mechanical tension of wires, those tension values are also reported in the Results tab. The fundamental frequency of the wire is identified by first computing a running cumulative sum of the data as a function of the stimulus frequency. This running sum is then passed to a Savitzky–Golay smoothing filter [24], using a sliding window size of about 3 Hz and a third-order polynomial, to perform a linear least-squares minimization to the data inside that window. The difference between the unfiltered and filtered data is taken such as to subtract the baseline on top of which the resonance lies. These processed amplitudes give rise to a unipolar resonance centered at the fundamental frequency of the wire over an essentially flat baseline. Both the raw and the processed amplitudes are plotted as a function of stimulus frequency in the Results tab for the eight DWA channels. The fundamental frequency is finally identified by determining at which frequency the maximum value of the processed amplitudes occurs. With the measurement of the wire's fundamental frequency, the associated tension can be computed from (1) since the software has knowledge of its length and linear density.

## VI. RESULTS

To investigate the performance of the DWA, a set of 17 parallel wires are laid down in an array across two PCBs that are secured to a frame 6 m apart of each other. The wires are made of a C17200 beryllium copper alloy and they have a 6-mil diameter and a linear density of $1.5 \times 10^{-4}$ kg/m, according to the nominal specifications of the manufacturer, similar to those used in [18]. The ends of the wires are soldered onto PCB pads that have a pitch of 4.8 mm. As they are being soldered, the wires are tensioned to 6.5 N via the use of a pulley and hanging test weights. All these wire properties match those of the DUNE wire arrays. Wires in the current case are expected to be at slightly different tensions due to the differences in handling and friction, which are not attempted to be rectified to show the capability of the DWA to measure different values simultaneously. The DWA is connected to the 17 wires via its flex cable and an adapter board that is attached to one of the two PCBs holding the wires. In addition, the wires are clamped with electrical tape onto an insulator 50 cm away from one of their ends to produce a node in the oscillation of the wires that corresponds to a fundamental frequency of about 200 Hz. This high-frequency region is chosen to illustrate the results of the DWA as it was identified by Garcia-Gamez et al. [18] to be at the limit of applicability of their implementation of the measurement method, given that the amplitude of the resonance decreases as the fundamental frequency increases.

A frequency sweep is performed around 200 Hz to measure the fundamental frequencies of eight of the wires. The DWA is configured with a frequency step size of 0.125 Hz, a stimulus duration before sampling of 1 s, a DAC output amplitude of 2048 mV, a digital potentiometer resistance of 53 kΩ for all the channels, one readout ac cycle of 32 samples, and dc high voltages of ±1.6 kV. The processed amplitudes are shown in Fig. 6 for the eight DWA readout channels. They correspond to the amplitudes shown within the GUI in Fig. 5. The measured fundamental frequencies span between 192.1 and 199.4 Hz. The fundamental frequencies are easily identifiable from these processed amplitudes even at these high frequencies. This indicates that the limit of applicability of the method has been extended compared with [18], especially considering that a larger wire pitch is used, 4.8 mm versus 3 mm, which further reduces the signal amplitude. This higher frequency limit is achieved by the DWA in part due to its higher dc stimulus voltages and in part due to its better noise mitigation. The specific high-frequency limit that can be measured depends on the wire pitch of the array and on the configuration of the instrument, since high voltages can be increased and the frequency step size can be decreased to better resolve resonances. Such a frequency limit can translate into a limit for measuring tension via (1).

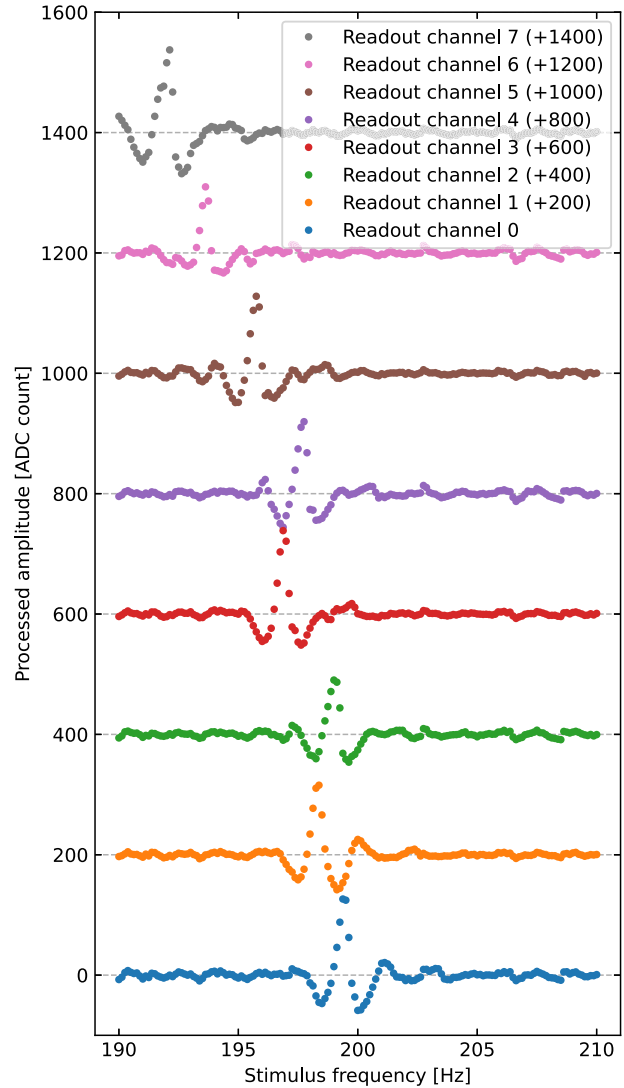

Fig. 6. Processed voltage amplitude of wires under test as a function of stimulus frequency for the eight DWA readout channels. The amplitudes are vertically offset for readability by a value specified in the legend.

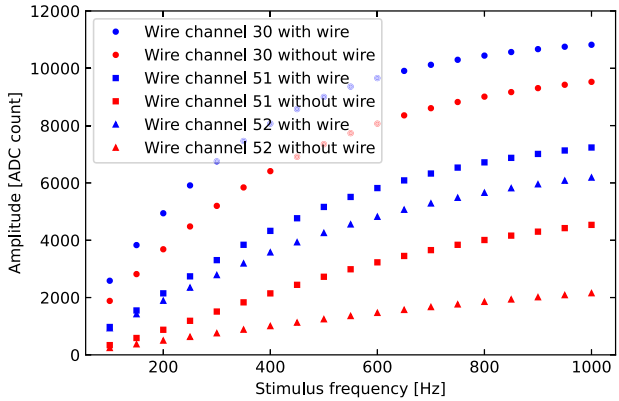

Fig. 7. Voltage amplitude as a function of stimulus frequency of selected wire channels with and without wires connected to them.

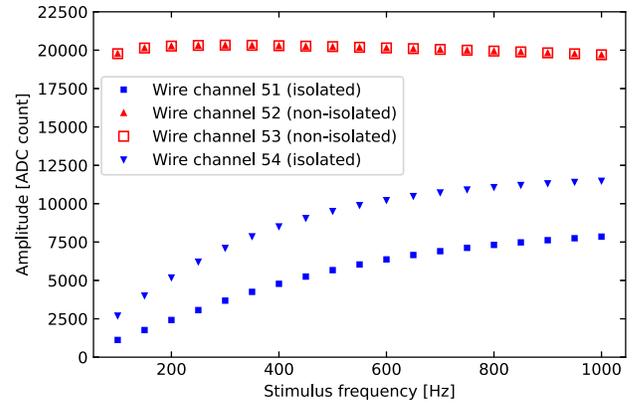

Fig. 8. Voltage amplitude as a function of stimulus frequency of consecutive isolated and nonisolated wires.

The accuracy of the measurement of fundamental frequencies by the DWA is verified by comparing the values obtained with it to those obtained with a laser–photodiode apparatus [3]. The apparatus measures the fundamental frequencies by shining a laser on a wire and measuring with a photodiode the modulation of the reflected light when the wire is strummed. A different set of wires with a length of 6 m and a tension of 6.5 N, corresponding to a fundamental frequency of about 17 Hz, are used for comparison. An agreement within 0.1 Hz with no significant bias between the fundamental frequencies measured by the DWA and the apparatus is obtained, translating to an accuracy of 1% in the tension measurement. Furthermore, a comparison of the laser–photodiode tension values to those directly measured by a force gauge is made by attaching a wire to the gauge while simultaneously shining the laser on it. The wires used in this case are 1 m long and are at a tension of either 5 or 6 N. The DWA is not used for this comparison due to its use of electrical signals on the wires, with which the gauge could interfere. Tension values measured with the laser–photodiode apparatus are systematically smaller by about 0.3 N than those measured with the gauge, corresponding to a bias of 5%. The size of this systematic error is compatible with a deviation in the value of the wire density within the manufacturing tolerance. A direct measurement of the wire density gives a value of $1.57 \times 10^{-4}$ kg/m. When using this measured density instead of the nominal one, the bias becomes negligible and the tension values measured with the laser–photodiode apparatus and the force gauge agree within 1%. This is consistent with the accuracy of the DWA as determined with different wire lengths and tensions. The accuracy of the DWA for tension measurement is thus determined to be at that 1% level, assuming the length and density of a wire are known to a better level than that. Although the determination of the DWA accuracy could be limited by the accuracy of the other instruments used in the comparisons, this accuracy level is comparable to that of the other measurement methods [1], [2], [3], [4], [5], [6], [7], [8], [9], [10], [11], [12], [13], [14], [15], [16], [17] and is thus not attempted to be characterized further. The DWA brings the accuracy of the method developed in [18] to the level of other methods while automating the measurement process.

Using the measured value of density, the wires of Fig. 6 have tensions measured to be in the 5.8–6.2-N range, with those limits corresponding, respectively, to channels 7 and 0. These measured values have an uncertainty of 0.3 N that is dominated by an uncertainty of 1 cm in the length of the wires, which is specific to the setup used to investigate this frequency region due to the clamping of wires with tape. The repeatability of the tension measurement is assessed by measuring the tension of the same wires multiple times. The repeated tensions are found to vary by less than 0.1%.

A frequency sweep is performed across most of the configurable frequency ranges to test the electrical continuity and isolation of 6-m-long wires. The minimum frequency of the sweep is 100 Hz, avoiding the need for MNS, while its maximum value is the maximum configurable value of 1 kHz. The other configuration parameters are a frequency step size of 50 Hz, a stimulus duration before sampling of about 0.2 s, a DAC output amplitude of 512 mV, a digital potentiometer resistance of 43 k$\Omega$ for all the channels, one readout ac cycle of 32 samples, and the dc high voltages are disabled. The voltage amplitudes obtained are shown in Figs. 7 and 8. In these plots, the effect of noise on the amplitude values, as determined by repeated sweeps, is smaller than the size of the data markers. Also, the shape of the data as a function of the stimulus frequency can be attributed to the high-pass $RC$ filter at the input of the VGAs.

The capacitance to which the high-pass filter is sensitive includes not only the contributions from the wires but also from the internal capacitance of the DWA itself. In particular, the internal capacitance depends on the routing through the switch matrix and the flex cable. Interchannel variations in internal capacitance and readout gain, which arise due to variations among the electronic components, result in different levels of rising baselines. As such, the wire channels of Fig. 7 are selected to show a representative range in the shape and size of the amplitudes. The figure shows data for the channels when they are connected to wires and when they are not. The differing size and shape of the amplitudes when a wire is present or not can be readily distinguished for a given channel in all the cases.

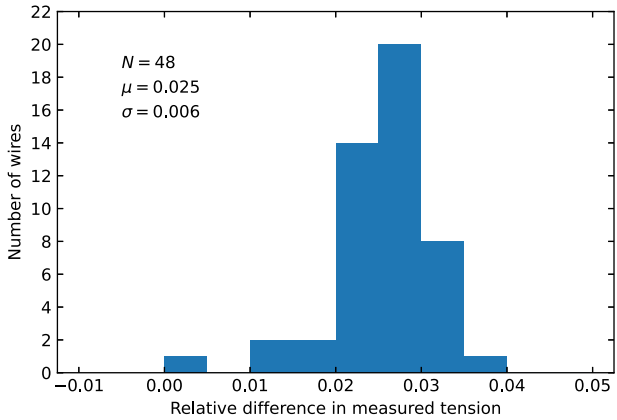

Fig. 9. Histogram of the relative difference in tension measurements between the DWA and laser–photodiode apparatus. The total number of wires measured $N$ and the mean $\mu$ and standard deviation $\sigma$ of the distribution are specified.

In Fig. 8, frequency sweeps are performed anew but with wire channels 52 and 53 brought into electrical contact through a jumper wire. It can be seen that the channels that are not isolated have larger amplitudes with a mostly flat shape as a function of the stimulus frequency, corresponding to reading out the ac stimulus voltage directly. Channels that are not isolated can easily be distinguished from when they are isolated based on if their amplitudes exceed a large threshold value. The DWA can therefore perform both the electrical continuity and isolation tests by comparing test data to per-channel reference values.

Environmental effects, such as variations in temperature and humidity, can have an impact on the amplitude of the signals digitized by the DWA. In the case of the tension measurement, since the method relies on identifying a relative change in signal amplitude as a function of stimulus frequencies, it is robust against such environmental effects. Similarly, it is also robust against variations in internal capacitance and readout gain among channels. In the case of the continuity and isolation tests, the impact of environmental effects on the outcome of the tests can be mitigated by recording reference values in the same environmental conditions as those present while conducting the tests.

The sweep in Fig. 6 takes about 3 min to complete, while those of Figs. 7 and 8 takes about 5 s. The main configuration parameters that determine the time taken for a sweep, other than the frequency range itself, are the frequency step size, the stimulus duration before sampling, and the number of readout ac cycles. The minimum value for that last parameter is used since it does not negatively impact the results. The other two parameters are constrained by the physics of the wire system for tension measurement sweeps. The step size cannot be increased arbitrarily as there is a risk of not resolving the resonance. If the stimulus duration is too short, the wire under test is not driven properly, resulting in a bipolar resonance that is not centered at the fundamental frequency due to the time needed for the wire under test to respond to the force. This frequency offset is investigated by performing frequency sweeps with a stimulus duration set to its minimum of 0 s and by measuring the same wires with the laser–photodiode apparatus. The relative difference between the tension measurements, i.e. $(T_{\text{DWA}} - T_{\text{laser}})/T_{\text{laser}}$, can be seen in Fig. 9. The positive mean value of the distribution illustrates the lag in the response of the wires with respect to the stimulus frequency. Calibration factors correcting for such bias in tension can be obtained from this type of distribution, here giving a value of 0.975, or by comparing resonance positions between sweeps with short and long stimulus durations. Calibrated sweeps with a short stimulus duration are found to be an order of magnitude faster than those that do not need a calibration while still being accurate. The spread in measurements of 0.6% is compatible with the previously stated 1% accuracy determined using a longer stimulus duration.

## VII. Conclusion

A DWA of mechanical tension and of electrical continuity and isolation has been designed, produced, and tested. Compared with the previous implementation of the measurement method of wire tension on which it is based, it can analyze a larger number of wires with less physical manipulation and it has a better signal-to-noise ratio, even for more challenging wire configurations, which results in a greater accuracy. It can also test electrical continuity and isolation.

Although the performance of the DWA as shown is good, more complicated wire arrays can offer some challenges, which could require modifications to what has been presented. For example, DUNE wire arrays have channels that have multiple independent wires connected in series. These wires can have different lengths, creating multiple resonances in a channel, possibly close to each other in frequency space. More complex software algorithms would therefore be needed to identify resonances and associate them to wires for those arrays.

The instrument can be improved in different ways. Faster sweeps that have a short stimulus duration can be calibrated, but ringing in the amplitude of readout voltages at frequencies that follow a resonance is observed. This can complicate the identification of resonances in channels that have multiple wires. To prevent any ringing from occurring, a stimulus voltage that is half a cycle out of phase could be applied on stimulus wires after voltage digitization is complete. This would forcefully damp the wire oscillations before moving on to the next stimulus frequency in the sweep. Another way to decrease the completion time of a sweep, which avoids the need to calibrate resonance position and any ringing, is to have a variable step size. A small step size is only useful near a resonance, such that a large step size could be used at first to find the general location of a resonance and then a smaller step size could be used to resolve it. Both these ideas would have to be implemented in the firmware to allow real-time adjustment of the hardware.

Finally, since frequency sweeps across the configurable frequency range are sensitive to the capacitance, the DWA could measure capacitance if it were calibrated with a known capacitive load. This would allow a measurement of wire capacitance, a quantity that is useful to know in particle detectors to take into account the crosstalk among wires in an array.

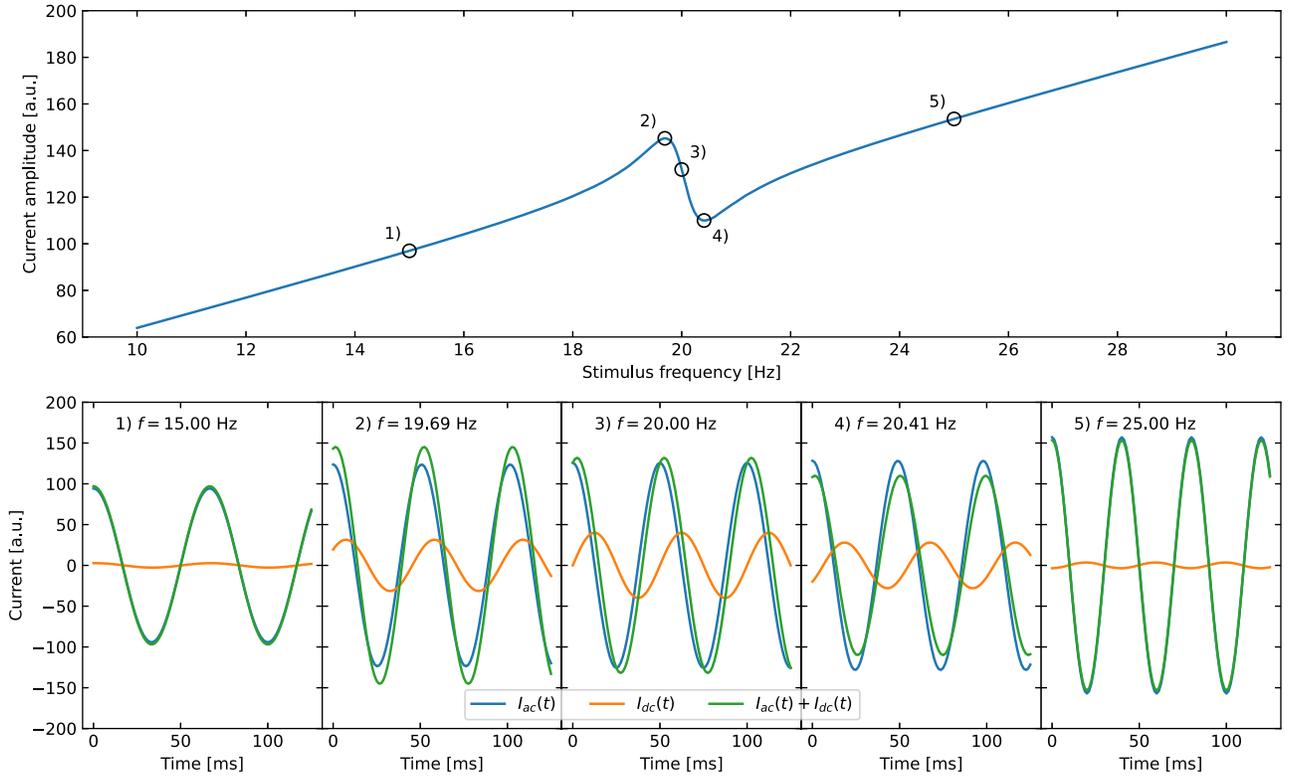

Fig. 10. Computed amplitude of the current in a wire under test as a function of stimulus frequency, in the top plot, and individual current contributions, and their sum, as a function of time for five selected stimulus frequencies, in the bottom plots. The five selected frequencies correspond to the five circled locations in the top plot, with corresponding number labels. The bottom plots show current values for the first 125 ms. Values of current are in arbitrary units (a.u.).

## APPENDIX

The oscillatory motion of a wire under test in the measurement method can be described by that of a usual driven and underdamped harmonic oscillator. Its amplitude $A(\omega)$ exhibits a unipolar resonance at a frequency near that of the fundamental frequency of the wire and is given by

$$A(\omega) = \frac{A}{\sqrt{(\omega_0^2 - \omega^2)^2 + (\gamma \omega)^2}} \quad (3)$$

where $A$ is a constant that is proportional to the electrical force on the wire and thus also to $V_{ac}$ and $V_{dc}$ of (2), $\gamma$ is a constant that is proportional to the damping coefficient of the wire, and $\omega_0 = 2\pi f_0$. The phase lag $\delta(\omega)$ of the oscillatory motion with respect to the driving force is given by

$$\delta(\omega) = \arctan \frac{\gamma \omega}{\omega_0^2 - \omega^2} \quad (4)$$

with values ranging between 0 and $\pi$, and with $\delta(\omega_0) = \pi/2$ in particular.

The capacitances $C_+$ and $C_-$ between the wire under test and its neighbors that are stimulated, respectively, with voltages $V_+$ and $V_-$ of (2) will oscillate around their at-rest value of $C$ as a result of the oscillatory motion of the wire under test. If $A(\omega)$ is much smaller than the wire pitch, these capacitance values as a function of time can be approximated by

$$C_\pm(t) = C\left(1 \pm a(\omega) \sin\left(\omega t - \delta(\omega)\right)\right) \quad (5)$$

where $a(\omega)$ is the amplitude of the relative change in capacitance, with $a(\omega) \ll 1$. This amplitude is proportional to $A(\omega)$ and as such also exhibits the same unipolar resonance. These time-varying capacitances and voltages result in a current in the wire under test that is

$$I(t) = \frac{d(C_+ V_+)}{dt} + \frac{d(C_- V_-)}{dt} \quad (6)$$

$$= 2C\omega\left(V_{ac}\cos \omega t + V_{dc} a(\omega) \cos\left(\omega t - \delta(\omega)\right)\right) \quad (7)$$

$$= I_{ac}(t) + I_{dc}(t) \quad (8)$$

$$= \omega a'(\omega) \cos\left(\omega t - \delta'(\omega)\right) \quad (9)$$

where (8) defines the symbols for the two terms in (7) and where the two primed quantities in (9) can be determined via trigonometric identities.

Given that $a(\omega)$ is small, $I_{dc}(t)$ is negligible, except near $\omega_0$ where its impact on the current amplitude can be observed. Due to $\delta(\omega)$, the two terms in (8) have a constructive (destructive) interference for values of $\omega$ approximately smaller (larger) than $\omega_0$. Therefore, the combined effect of $a(\omega)$ and $\delta(\omega)$ on $a'(\omega)$ results in it having an essentially constant value throughout the frequency range, except near $\omega_0$ where a bipolar resonance centered approximately at $\omega_0$ occurs. The additional factor of $\omega$ in (9) brings a rising baseline to the current amplitude.

To illustrate the effect of the interference between the two terms of (8), Fig. 10 shows examples of the two currents and of their sum for various stimulus frequencies in the bottom plots, with the amplitude of the current sum reported in the top plot. In this illustration, arbitrary numerical values have been assigned to the various parameters but still in a way that is representative of what could be measured in a wire array, in particular $f_0 = 20$ Hz. It can be seen that the current in the wire under test is dominated by $I_{ac}(t)$, but that the constructive and destructive interferences between $I_{ac}(t)$ and $I_{dc}(t)$ have an observable effect near the fundamental frequency. The effect is observable not only on the amplitude of the current sum but also on its phase. Indeed, a mechanism similar to the one that produces a bipolar resonance in amplitude also produces a unipolar resonance in $\delta'(\omega)$. The resonance in the phase of the current is not attempted to be used as a measurement method however. It can also be seen from the figure that the center of the resonance is slightly offset relative to the fundamental frequency of the wire. No attempt is made to correct for such offsets in measurements.

## Acknowledgment

The authors thank the DUNE Collaboration for their interest in this work, Garcia-Gamez et al. for helpful discussions about their work, the Physical Sciences Laboratory of University of Wisconsin–Madison for graciously providing material and technical support, and Prof. M. Soderberg of Syracuse University for sharing his laser–photodiode apparatus.

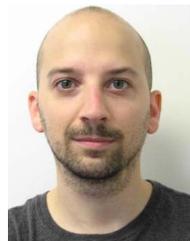

**Sebastien Prince** received the Ph.D. degree in experimental high-energy physics from McGill University, Montreal, QC, Canada, in 2018, for work done on the ATLAS experiment.

Since then, he has been an NSERC and FRQNT Post-Doctoral Fellow with the Department of Physics, Harvard University, Cambridge, MA, USA, working on experimental neutrino physics as a member of the DUNE and MicroBooNE Collaborations.

Dr. Prince is certified as a Professional Physicist by the Canadian Association of Physicists. He is also a member of the American Physical Society.

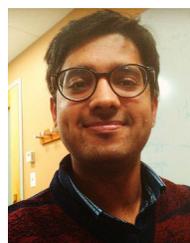

**Pratyush Anand** received the B.Tech. degree in engineering physics from IIT Madras, Chennai, India, in 2019. He is currently pursuing the master's degree with the Department of Physics, ETH Zurich, Zurich, Switzerland.

He is currently a Research Fellow with the Department of Physics, Harvard University, Cambridge, MA, USA, in the field of quantum optics, quantum sensing, and quantum networks. He has also worked in high-energy physics on the ATLAS, BESIII, DUNE, MicroBooNE, and muEDM experiments.

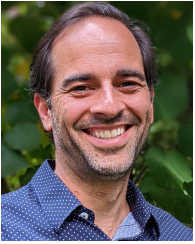

**James Battat** received the Ph.D. degree in astronomy from Harvard University, Cambridge, MA, USA, in 2008, from using lunar laser ranging to carry out tests of gravity and other fundamental physics.

He was a Pappalardo Fellow in physics with the Massachusetts Institute of Technology, Cambridge. He is currently an Associate Professor of physics and a Chair of the Physics Department, Wellesley College, Wellesley, MA. He was a member of the DMTPC and DRIFT Collaborations and is currently a member of the DUNE Collaboration. His work has been recognized with a Sloan Foundation Fellowship.

Prof. Battat is a member of the American Physical Society and the American Astronomical Society.

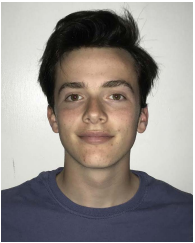

**Russell Farnsworth** is currently pursuing the A.B. degree in physics with Harvard University, Cambridge, MA, USA.

He has been an Undergraduate Research Assistant with Harvard University, where he has been working on the DUNE and ATLAS experiments. His research interests include neutrino physics and high-energy physics.

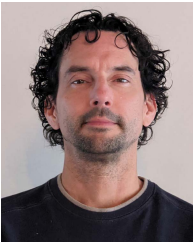

**Nathan Felt** received the B.S. degree in electrical and computer engineering from Western Michigan University, Kalamazoo, MI, USA, in 1997.

Since then, he has been an Electrical Engineer with the Department of Physics, Harvard University, Cambridge, MA, USA. He has worked on detectors of experiments in high-energy physics, including CDF, MINOS, NOvA, ATLAS, and DUNE.

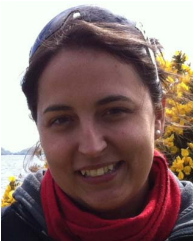

**Roxanne Guenette** received the Ph.D. degree in high-energy astrophysics from McGill University, Montreal, QC, Canada, in 2010, for work done on the VERITAS experiment.

From 2010 to 2013, she was a Post-Doctoral Associate with the Department of Physics, Yale University, New Haven, CT, USA. From 2013 to 2017, she was an STFC Ernest Rutherford Fellow with the Department of Physics, University of Oxford, Oxford, U.K. From 2017 to 2022, she was an Assistant Professor with the Department of Physics, Harvard University, Cambridge, MA, USA. Since 2022, she has been a Professor of particle physics with the Department of Physics and Astronomy, University of Manchester, Manchester, U.K.

Dr. Guenette is a member of the DUNE, MicroBooNE, SBND, and NEXT Collaborations.

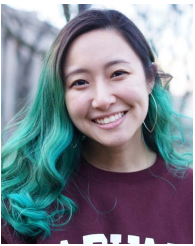

**Shion Kubota** received the B.A. degree in physics from the Mount Holyoke College, South Hadley, MA, USA, in 2019. She is currently pursuing the Ph.D. degree with the Department of Physics, Harvard University, Cambridge, MA, working as a member of the DUNE and Q-Pix Collaborations.

She has also worked on the ATLAS experiment. She is a Scholar of the Ezoe Memorial Recruit Foundation and the Masason Foundation.

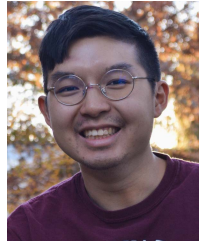

**Austin Li** is currently pursuing the A.B. degree in physics and computer science with Harvard University, Cambridge, MA, USA, in 2023.

He has been an Undergraduate Research Fellow with Harvard University, where he is working on the DUNE experiment. His research interests include particle physics and quantum computing.

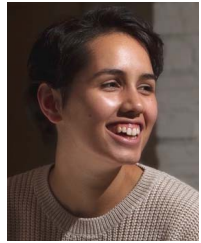

**Em Murdock** is expected to receive the A.B. degree in physics, environmental science, and public policy with Harvard University, Cambridge, MA, USA, in 2023.

He worked as a Research Assistant in neutrino physics with Harvard University. His research interests include environmental science and policy, with published works relating to climate-oriented stimulus policies in the wake of the COVID-19 pandemic.

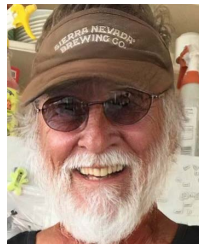

**John Oliver** received the Ph.D. degree in theoretical high-energy physics from Boston University, Boston, MA, USA, in 1974.

He worked at P. R. Mallory and Company Inc., Burlington, MA, USA, and GCA Corporation, Bedford, MA, USA. Since 1980, he has been an Electrical Engineer with the Department of Physics, Harvard University, Cambridge, MA. He officially retired in 2011 but has continued to work part-time since then. He has worked on detectors of experiments in high-energy physics and on astronomical instrumentation, including CDF, CLEO, MINOS, NOvA, ATLAS, LSST, DUNE, and NEXT.

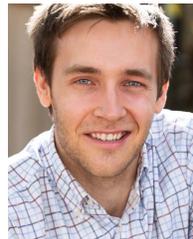

**Chris Stanford** received the Ph.D. degree in physics from Princeton University, Princeton, NJ, USA, in 2017, as a member of the DarkSide Collaboration.

He was a KIPAC Fellow with Stanford University, Stanford, CA, USA, and a member of the SuperCDMS Collaboration. He is currently a Post-Doctoral Fellow with the Department of Physics, Harvard University, Cambridge, MA, USA, and a member of the DUNE and NEXT Collaborations. Throughout his career, his focus has been on instrumentation and detector Research and Development for experiments in particle astrophysics.

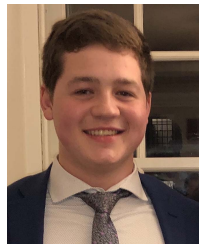

**Jackson Weaver** received the A.B. degree in physics and mathematics from Harvard University, Cambridge, MA, USA, in 2021. He is currently pursuing the master's degree with the Department of Materials Science and Metallurgy, University of Cambridge, Cambridge, U.K., working on simulations of incommensurate crystals.

He has also worked on the DUNE experiment.

Mr. Weaver is a member of the American Physical Society.